\documentclass[ twocolumn,
		        aps,
		        superscriptaddress,
		        longbibliography,
		        pra
		       ]{revtex4-1}

\usepackage{hyperref}
\usepackage{graphicx}
\usepackage{wrapfig}
\usepackage{booktabs}
\usepackage{amsmath}
\usepackage{amssymb}
\usepackage{braket}
\usepackage[dvipsnames]{xcolor}
\usepackage{ulem}

\begin{document}
\title{Conditional nonclassical field generation in cavity QED}

\author{Karsten Weiher}
\affiliation{Institut f\"ur Physik, Universit\"at Rostock, D-18051 Rostock, Germany}

\author{Elizabeth Agudelo}\email{elizabeth.agudelo@oeaw.ac.at}
\affiliation{Institute for Quantum Optics and Quantum Information (IQOQI) Vienna, Austrian Academy of Sciences, Boltzmanngasse 3, 1090 Vienna, Austria}

\author{Martin Bohmann}\email{martin.bohmann@ino.it}
\affiliation{Quantum Science and Technology in Arcetri (QSTAR),
Istituto Nazionale di Ottica-Consiglio Nazionale delle Ricerche (INO-CNR),
and European Laboratory for Nonlinear Spectroscopy (LENS), Largo Enrico Fermi 2, I-50125 Firenze, Italy}

\begin{abstract}
    We introduce a method for the conditional generation of nonclassical states of light in a cavity.
    We consider two-level atoms traveling along the transverse direction to the cavity axis and show that by conditioning on one of the output measurements 
    nonclassical field states are generated.
    The two-level atoms are prepared in the ground state and we conditioned them on the events in which they are also detected in the ground state.
    Nonclassical properties of the cavity mode are identified and characterized.
    This includes: quadrature squeezing, sub-Poissonian photon-number distributions, and negative Wigner functions.
    We determine the optimal parameter regions where the corresponding nonclassical features are most distinct.
\end{abstract}
\maketitle

\section{Introduction}

    The generation and verification of nonclassical states are key tasks in quantum optics and quantum information.
    Besides their fundamental role in the understanding of quantum effects and correlations, the preparation and identification of genuine quantum features are becoming increasingly important as they are necessary for applications in quantum technologies such as communication \cite{Gisin_2007, Ralph_1999}, metrology \cite{Paris_2009, Safranek_2015}, and computation \cite{nielsen_2011, Lloyd_1999, Menicucci_2006, Gu_2009}.
    In particular, continuous-variable nonclassicality has been identified as a resource for quantum technologies \cite{yadin_2018, kwon_2019}. 
    In the context of these resource theories, entanglement may be considered as a secondary effect, as it is easily obtained by nonclassical states and passive linear optics.
    Notably, a given amount of nonclassicality is fully equivalent, as a resource, to exactly the same amount of entanglement (when the quantification is based on the quantum superposition principle) \cite{vogel_2014}.
    Therefore, it is crucial to develop efficient strategies for the preparation of nonclassical states. 
 
    An ideal platform for engineering and studying nonclassical states is cavity quantum electrodynamics (QED) \cite{raimond_2001,haroche_2006,walther_2006}.
    In cavity QED one investigates the interaction of a radiation field inside a cavity with atoms and the resulting transfer of quantum information from matter to radiation and vice versa.
    The basic interactions and resulting dynamics is described via the Jaynes-Cummings Hamiltonian \cite{jaynes_1963}.
    Different effects have been characterized using this model, such as Rabi oscillations \cite{haroche_1985,gallas_1985}, the collapse and revival of probabilities \cite{rempe_1987}, or its nonclassical correlations \cite{hangley_1997}.
    Other nonclassical effects such as quantum jumps have been also observed in a cavity \cite{gleyzes_2007}.
    
    In particular, it is possible to generate nonclassical states of the cavity field, by a careful control of the light-atom interaction.
    This includes the preparation of sub-Poissonian photon statistics \cite{rempe_1990}, photon-number states \cite{varcoe_2000,brattke_2001,sayrin_2011,bertet_2002}, squeezed states \cite{raizen_1987}, or superposition states such as Schr\"odinger-cat state \cite{brune_1996,auffeves_2003}.
    
    In cavity QED, several strategies for the conditional-state preparation have been introduced.
    Among them is the generation of Fock states by the conditional addition of photons to a cavity vacuum state \cite{krause_1987,varcoe_2000}.
    In this case one uses the fact that an atom in the excited state can add a photon to the cavity field.
    Other approaches use adaptive measurements on the atomic state for enhancing squeezing \cite{gerry_1997,chang-qi_1992}, for hole burning in the Fock space \cite{malbouisson_2001,avelar_2005}, or photon amplification \cite{moya-cessa_1994,moya-cessa_1999}.
    More general strategies for the conditional quantum-state engineering of the cavity field have been also discussed \cite{vogel_1993,garraway_1994,kozhekin_1996,ghosh_1997}.

    In this paper, we introduce a method for a conditional generation of nonclassical cavity field states based on post-selection.
    In particular, we consider the situation where the two-level atoms enter and exit the cavity (which is prepared in a coherent state) in its ground state.
    Although, in this case, in the end no photons are added to or subtracted from the initial cavity field, the obtained field state features various nonclassical properties.
    We derive the corresponding conditional field state in the photon-number basis and calculate the success probability of the protocol.
    We present and analyze different nonclassical characteristics of the generated quantum state.
    This includes quadrature squeezing, negative values of the Mandel parameter, and negativities of its Wigner function.
    Optimal parameter regions for these different quantum features are identified.
    Importantly, the proposed protocol is applicable to scenarios with a rather low atom-field coupling parameter which assures the applicability to many experimental realisations. 
    Thus, our approach presents an easily implementable method for the generation of nonclassical cavity fields.
    
    The paper is structured as follows.
    In Sec. \ref{sec:protocol}, we introduce the conditional state-preparation protocol and the corresponding quantum state of the cavity field.
    In Sec. \ref{sec:nonclassicality}, we study and characterize the nonclassicality features of the generated conditional state.
    The results are discussed and compared in Sec. \ref{sec:discussion}.
    We summarize the findings and conclude in Sec \ref{sec:conclusions}.
   
\section{Conditional nonclassical state generation}
\label{sec:protocol}
    In this section, we present the protocol for the conditional generation of nonclassical cavity-field states.
    First, we introduce the considered setup and motivate the protocol.
    Second, we describe the proposed protocol mathematically and derive the conditional quantum state of the cavity field.
    Finally, we calculate the success probability of the adaptive strategy.

\subsection{Setup and motivation}
    The physical scenario we are interested in is a QED system which consists of two-level Rydberg atoms passing through a high-quality microwave cavity.
    We consider the standard configuration in cavity QED \cite{raimond_2001}, where ground and excited states are related to Rydberg states with $n=50$ and $n=51$, respectively.
    As initial conditions, we have a two-level atom prepared in its ground state and a cavity field initially prepared in a (classical) coherent state.
    We consider the case in which the frequency of the cavity mode is resonant with the atomic transition.
    In this case, the atom-cavity interaction can be described via the resonant Jaynes-Cummings model \cite{jaynes_1963} in the strong coupling regime.
    The corresponding interaction Hamiltonian in the rotating-wave approximation reads
     \begin{align}
        \hat{H}^{(I)} = \hbar\Omega_0\left( \hat{\sigma}_+\hat{a} + \hat{\sigma}_-\hat{a}^\dag \right)/2,
        \label{eq:Ham}
    \end{align}
    with the atomic transition operators $\hat{\sigma}_+$ and $\hat{\sigma}_-$ and the photon annihilation and creation operators $\hat{a}$ and $\hat{a}^\dag$, respectively.
    The coupling constant $\Omega_0$ describes the interaction strength between the cavity field and the atoms which depends on the systems characteristics such as the dipole transition between the states $\ket{g}$ and $\ket{e}$ and the amplitude of the cavity field.

\subsection{Protocol} 
    We analyze the case in which the atoms enter the cavity in the ground state, i.e.,  $\hat{\rho}_{A}(0) = \ket{g}\bra{g}$, and the cavity field is initially prepared in a coherent state, $\hat{\rho}_{F}(0)= \ket{\alpha}\bra{\alpha}$.
    The composite atom-field system is described by its density operator at $t= 0$,
    \begin{align*}
        \hat{\rho}(0) = \hat{\rho}_{\mathrm{F}}(0)\otimes \hat{\rho}_{\mathrm{A}}(0) =
        \begin{pmatrix}
	    0 & 0 \\
	     0 & \hat{\rho}_{\mathrm{F}}(0)
	    \end{pmatrix} .
    \end{align*}
    Here the matrix expansion of $\hat \rho$ is given in the basis of the atomic states.
    The system evolves according to the interaction Hamiltonian in Eq. \eqref{eq:Ham} to 
    \begin{align*}
        \hat{\rho}(t) &=  \mathrm{e}^{-i\hat{H}^{(I)}t/\hbar}\,\hat{\rho}(0)\,\mathrm{e}^{i\hat{H}^{(I)}t/\hbar}, 
    \end{align*}
    where the matrix elements are explicitly
    \begin{align*}
        \hat{\rho}_{11}(t)&=-\hat{\mathrm{S}}'\hat{\rho}_{\mathrm{F}}(0)\hat{\mathrm{S}},  \\ \hat{\rho}_{12}(t)&=\hat{\mathrm{S}}'\hat{\rho}_{\mathrm{F}}(0)\hat{\mathrm{C}}', \\
	    \hat{\rho}_{21}(t)&=-\hat{\mathrm{C}}'\hat{\rho}_{\mathrm{F}}(0)\hat{\mathrm{S}}, \\ \hat{\rho}_{22}(t)&=\hat{\mathrm{C}}'\hat{\rho}_{\mathrm{F}}(0)\hat{\mathrm{C}}'.
    \end{align*}
    The operators $\hat{\mathrm{S}}'$ , $\hat{\mathrm{S}}$ and $\hat{\mathrm{C}}'$ are defined as
    \begin{align*}
        \hat{\mathrm{S}} &= -i\hat{a}^\dag\frac{\sin\left(\frac{\Omega_0}{2} t\sqrt{\hat{a}\hat{a}^\dag}\right)}{\sqrt{\hat{a}\hat{a}^\dag}}, \\
        \hat{\mathrm{S}}'&= -i\hat{a}\frac{\sin\left(\frac{\Omega_0}{2} t\sqrt{\hat{a}^\dag\hat{a}}\right)}
        {\sqrt{\hat{a}^\dag\hat{a}}},  \\
        \hat{\mathrm{C}}'&= \cos\left( \frac{\Omega_0}{2} t\sqrt{\hat{a}^\dag\hat{a}} \right).
    \end{align*}
    These operators describe the time evolution of the atom-field system.
    The first element, $\hat{\rho}_{11}(t)$ corresponds to the atom absorbing one photon of the cavity field, while $\hat{\rho}_{22}(t)$ corresponds to the atom staying in the ground state.
    
    Let us introduce the coupling parameter $r = \Omega_0 t/2$ which is the relevant parameter in the description of the interaction between the cavity field and the atoms.
    The operators $\hat{\rho}_{11}(r)$ and $\hat{\rho}_{22}(r)$ are given by
    \begin{gather*}
        \hat{\rho}_{11}(r) = \mathrm{e}^{-\left|\alpha\right|^2} \sum_{n,m}  \mathrm{c}'_{nm} (\alpha) \ket{n - 1}\bra{m - 1}, \\[1em]
       \text{with} \quad \mathrm{c}'_{nm} (\alpha) = \frac{\alpha^n\alpha^{*m}}{\sqrt{n!m!}}\sin\left(r \sqrt{n}\right) \sin\left(r\sqrt{m} \right)
    \end{gather*}
    and
     \begin{gather*}
        \hat{\rho}_{22}(r) = \mathrm{e}^{-\left|\alpha\right|^2} \sum_{n,m} \mathrm{c}_{nm} (\alpha)\ket{n}\bra{m}, \\[1em]
       \text{with} \quad \mathrm{c}_{nm} (\alpha) = \frac{\alpha^n\alpha^{*m}}{\sqrt{n!m!}}\cos\left(r \sqrt{n}\right) \cos\left(r \sqrt{m} \right),
    \end{gather*}
    respectively.
    We focus on the field state which is obtained by post-selecting the atom to be in the ground state after the interaction with the cavity field. 
    This state is expressed through projecting $\hat\rho(r)$ onto the atomic ground state and re-normalizing the corresponding field density operator
    \begin{align}
        \hat{\rho}_{\mathrm{ps}}(r) &= \frac{\hat{\rho}_{22}(r)}{\mathrm{Tr}[\hat{\rho}_{22}(r)]}.
        \label{rho_ps}
    \end{align}
    The state $\hat{\rho}_{\mathrm{ps}}(r)$ relates to the conditional field-state generation by post-selection (PS) using one atom.
    
    We can further generalize this approach to the subsequent interactions and PS with $N$ atoms.
    We assume that all atoms have the same properties and interact for the same time with the cavity field, i.e., the coupling parameter $r$ is equal for all atoms. 
    Note that, it is possible to tune the coupling parameter through the control of the interaction time, for example, using a stark shift \cite{haroche_2006}.
    In the case of post-selecting on all  $N$ atoms being measured in the ground state after the interaction with the cavity field, we obtain the PS density operator $\hat{\rho}^{N}_{\mathrm{ps}}(r)$.
    To calculate $\hat{\rho}^{N}_{\mathrm{ps}}(r)$, we successively apply the time evolution of each atom with the corresponding cavity-field state and subsequently project on the atomic ground state.
    Thus, the $N$-atoms PS density operator $\hat{\rho}^{N}_{\mathrm{ps}}(r)$ is given by
    \begin{align*}
        \hat{\rho}^{N}_{\mathrm{ps}}(r) &= \dfrac{\hat{\rho}^{N}_{22}(r)}{\mathrm{Tr}\left[\hat{\rho}^{N}_{22}(r)\right]} ,
    \end{align*}
    which can be calculated through the iterative operation
    \begin{align*}
        \hat{\rho}^{N}_{22}(r) &= {\hat{\mathrm{C}}}'\dfrac{\hat{\rho}^{N-1}_{22}(r)}{\mathrm{Tr}\left[\hat{\rho}^{N-1}_{22}(r)\right]}{\hat{\mathrm{C}}}',
    \end{align*}
    starting from the one-atom PS state $\hat{\rho}^{1}_{22}(r)=\hat{\rho}_{22}(r)$.
    Applying this procedure, eventually, yields
    \begin{equation}\label{eq:statePSN}
        \hat{\rho}^{N}_{\mathrm{ps}}(r) =  \frac{\sum_{n,m} \mathrm{c}_{nm} (\alpha,N)\ket{n}\bra{m}}
        {\sum_{n}\mathrm{c}_{nn} (\alpha,N)},
    \end{equation}
    with
    \begin{equation*}
        \mathrm{c}_{nm} (\alpha,N) =  \dfrac{\alpha^n\alpha^{*m}}{\sqrt{n!m!}}\cos^{N}\left(r \sqrt{n}\right) \cos^{N}\left(r \sqrt{m} \right).
    \end{equation*}
    
    We are interested in identifying the nonclassicality properties of the state in Eq. \eqref{eq:statePSN} which depends on the coupling parameter $r$, the number of atoms $N$, and the amplitude of the initial coherent state $\alpha$.
    In our consideration, we will focus on the parameter range of $0\leq r \leq 3 $ and up to five PS atoms. 
    If we assume a typical vacuum Rabi frequency of $\Omega_0=314$~kHz \cite{raimond_2001,haroche_2006}, a coupling parameter $r=3$ corresponds to an effective atom-cavity interaction time of $19\,\mu\mathrm{s}$.
    These time scales are short compared to typical cavity damping times which are of the order of  milliseconds.
    Note that, to consider non-interacting successive atoms, there is a limit distance between them which is of the order of few micrometers.
    Considering typical atomic velocities (c.f. \cite{raimond_2001,haroche_2006}) the necessary time gap between successive atoms can be estimated to be $7$~ns. 
    This assures the feasibility of the introduced state-preparation protocol even for several atoms, and allows us to neglect decoherence effects.
    In the following, we will fix the initial cavity-field state to be a coherent state with coherent amplitude $\alpha=\sqrt{10}$.
    The obtained results are qualitatively similar for different initial coherent amplitudes.
    We note that our analytical treatment allows for a general description with arbitrary $\alpha$.
    Before discussing the nonclassical properties, we will calculate the success probability of the post-selection.
    
\subsection{Success probability}
    To prepare the cavity field in the state described by the density operator in Eq. \eqref{eq:statePSN}, we have to successfully post-select all $N$ atoms.
    The overall success probability for $N$ atoms is, thus, the product of the success probability of each individual atom, 
    \begin{equation*}
        P_N = \prod_{i = 0}^N p_i.
    \end{equation*}
    Here, $p_i$ corresponds to the probability of finding the $i$-th atom in the ground state after it passed the cavity provided that also all atoms before had been detected in the ground state. 
    This probability is obtained via tracing over the field states in the $i$th-atom density operator and projecting to the ground state of the atom $p_i = \bra{g}\mathrm{Tr}\left[\hat{\rho}^{i}(r)\right]\ket{g}$.
    
    For one atom, we easily calculate the PS probability $P_1 = \mathrm{Tr}\left[\hat{\rho}^1_{22}(r) \right]$. 
    To calculate $P_2$ it is vital to take into account the normalization of the density operator after the first PS, given in Eq. \eqref{rho_ps}.
    This then yields
    \begin{align*}
        P_2 = p_1 p_2=\mathrm{Tr}\left[\hat{\rho}^1_{22}(r) \right]  \dfrac{\mathrm{Tr}\left[\hat{\rho}^2_{22}(r) \right]}
        {\mathrm{Tr}\left[\hat{\rho}^1_{22}(r) \right]}
        = \mathrm{Tr}\left[\hat{\rho}^2_{22}(r) \right].
    \end{align*}
    Similarly, by repeating this procedure, we find the $N$-atom success probability to be
    \begin{align*}
        P_N = \mathrm{Tr}\left[\hat{\rho}^N_{22}(r) \right].
    \end{align*}
    
    \begin{figure}[ht]
        \center
        \includegraphics[width = \columnwidth]{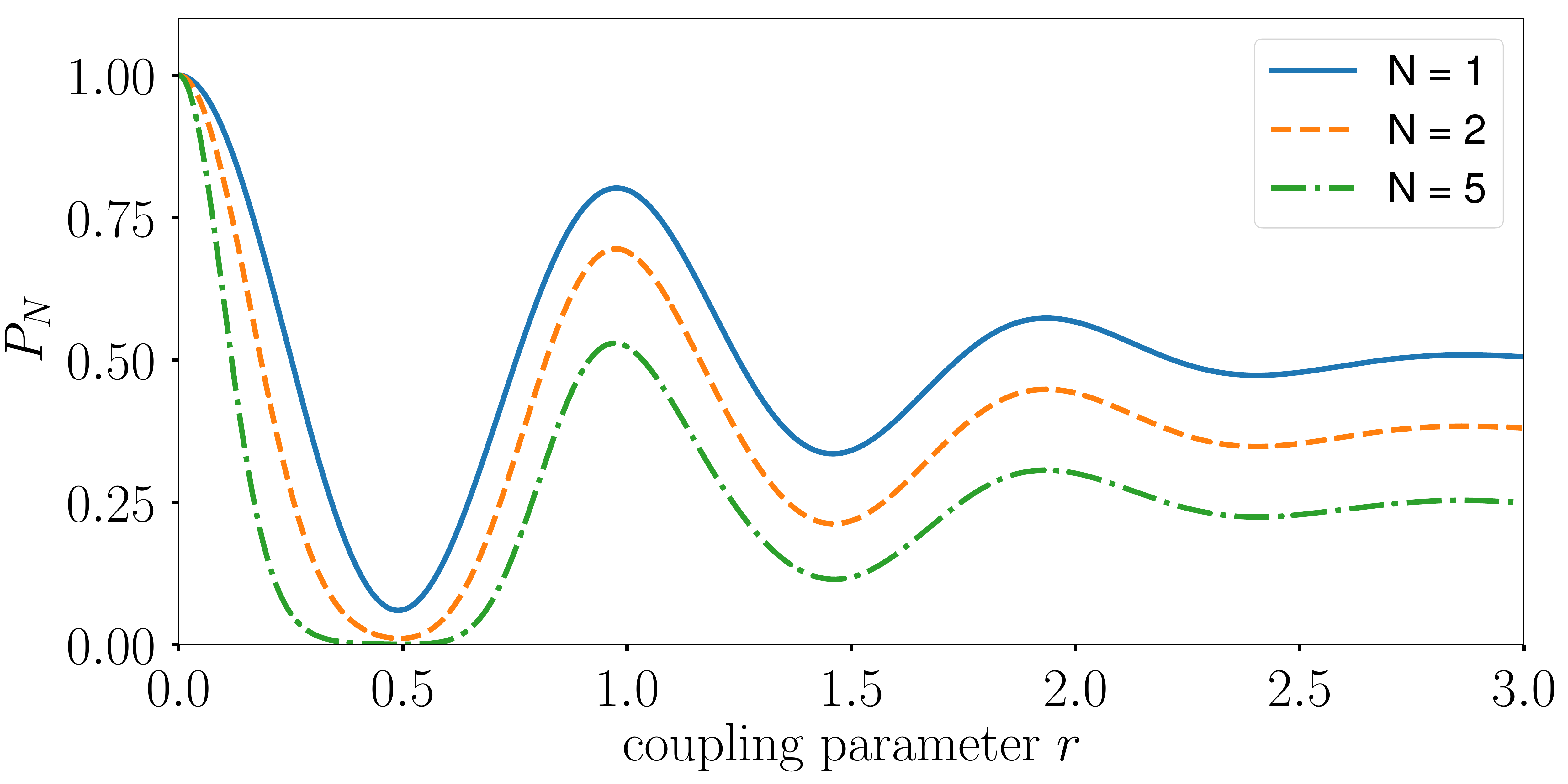}
        \caption{
        Variation of the success probability $P_N$ of the PS state in terms of the coupling parameter $r$ for $N=1$ (blue, solid), $N=2$ (orange, dashed), and $N=5$ (green, dash-dotted).
        }
        \label{fig:success}
    \end{figure}
    
    The success probability of the conditional field-generation protocol is shown in Fig. \ref{fig:success} for three different numbers of atoms.
    We observe that $P_N$ oscillates with respect to the coupling parameter $r$.
    In particular, for one atom it is exactly given by the probability of observing the atom in the ground state after it passes the cavity, which is nothing else than its Rabi oscillation.
    With more atoms, the success probability decreases faster while the oscillatory behavior of the one-atom case is preserved.
    Overall, the success probabilities are rather high (above 10\%, except for $r\approx 0.5$) which guarantees the applicability of the protocol in reasonable experimental times.
    Furthermore, the probability of observing $N$ atoms in the ground state is higher than the $N$-th power of the probability of the first atom, i.e., $P_{\mathrm{N}}>p_1^N$.
    Accordingly, finding the first atom in the ground state, in general, increases the likelihood of finding the subsequent ones in the ground state as well. 
    
    In the following, we will examine the nonclassical properties of the obtained cavity-field state for $N = 1,2$, and $5$ atoms. 
    For this number of atoms pronounced nonclassical features can be observed.
    Higher values of $N$ do not improve the observed features significantly or produce new effects.

\section{Nonclassical features}
\label{sec:nonclassicality}
    In this section, we investigate various aspects of the conditionally generated state of the cavity field.
    This includes quadrature squeezing, sub-Poissonian photon-number distributions characterized by the Mandel $Q_{\mathrm{M}}$ parameter, and negativities of the Wigner phase-space distribution.
    We discuss the variations of these properties with the coupling parameter and the number of PS atoms.

\subsection{Quadrature squeezing}
    Quadrature squeezing is the suppression of the quadrature noise below the vacuum noise level.
    The quadrature operator is defined in terms of the phase parameter $\varphi$ as $\hat x(\varphi) =\hat{a}\mathrm{e}^{-i\varphi} + \hat{a}^\dag \mathrm{e}^{i\varphi}$.
    A quantum state shows quadrature squeezing if its variance fulfills 
    $\langle (\Delta \hat x(\varphi))^2\rangle < 1.$
    Consequently, squeezing can be identified through the following condition:
    \begin{align}\label{eq:sqeezing}
        \braket{{:}\left(\Delta \hat x(\varphi)\right)^2{:}}=\braket{\left(\Delta \hat x(\varphi)\right)^2} - 1 <0,
    \end{align}
    where ${:}\quad{:}$ denotes the normal-order prescription (see, e.g., \cite{vogel_2006}).
    Note that, additionally, any proper quantum state needs to fulfill the uncertainty relation $\braket{(\Delta \hat x(\varphi))^2}\braket{(\Delta \hat x(\varphi + \pi/2))^2} \geq 1 $.
    The first and second moments of the quadrature operator for the $N$-atom PS state are
      \begin{widetext}
      \begin{gather*}
         \braket{\hat x(\varphi)} = \frac{2\mathrm{Re}\left(\alpha \mathrm{e}^{-i\varphi}\right) \mathrm{e}^{-|\alpha|^2}}{\mathrm{Tr}\left[ \hat{\rho}^{N}_{22} \right]} \sum_{n} \frac{|\alpha|^{2n}}{n!}\cos^{N}\left(r \sqrt{n}\right) \cos^{N}\left(r \sqrt{n + 1} \right),\quad\text{and}\\
        \braket{\hat x(\varphi)^2} = \frac{\mathrm{e}^{-|\alpha|^2}}{\mathrm{Tr}\left[ \hat{\rho}^{N}_{22} \right]} \sum_{n,m} \frac{|\alpha|^{2n}}{n!}
        \Big[ 2\mathrm{Re}\left(\alpha^2\mathrm{e}^{-i2\varphi}\right)  \cos^{N}\left(r\sqrt{n}\right) \cos^{N}\left(r \sqrt{n+2} \right) 
         + 2|\alpha|^2\cos^{2N}\left(r \sqrt{n + 1} \right) + \cos^{2N}\left(r\sqrt{n} \right) \Big],
    \end{gather*}
    \end{widetext}
    respectively.
    
    \begin{figure}[ht]
        \center
        \includegraphics[width = \columnwidth]{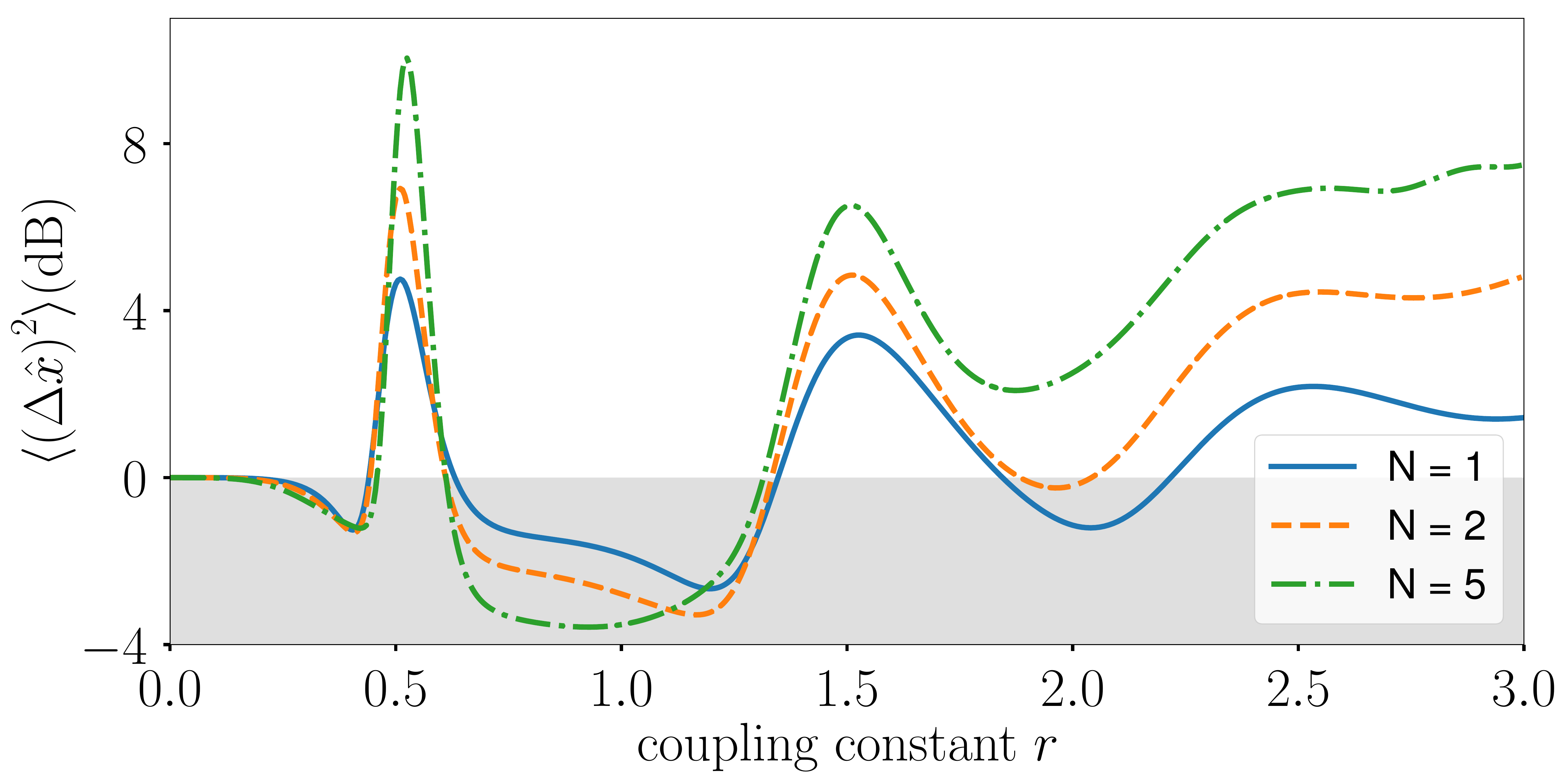} 
        \caption{
         Variations on the quadrature squeezing with the coupling parameter $r$ for the $N$-atom state with one, two, and five atoms and $\varphi=0$.
         To show squeezing in dB, we considered $10\log_{10}\left(\braket{(\Delta \hat x(\varphi=0))^2}/\braket{(\Delta \hat x)^2}_\textrm{vac}\right)$.
         A quantum state shows quadrature squeezing in dB if its variance fulfills  $\braket{(\Delta \hat x(\varphi))^2}(\textrm{dB}) < 0$, which corresponds to condition \eqref{eq:sqeezing}.
         For different sets of parameters, all studied states exhibit quadrature squeezing, which is indicated through the shaded area.
         }
        \label{fig:QuadSquee}
    \end{figure}
    
    In Fig.~\ref{fig:QuadSquee}, the variance of the quadrature operator in dB of the generated state is shown for different numbers of PS atoms in dependence on the coupling parameter $r$.
    We note that the considered state is most squeezed along the amplitude quadrature ($\varphi=0$). 
    This can also be seen for the Wigner function representation in Sec. \ref{sec:Wigner}.
    Therefore, we only analyze and show squeezing along the $\hat x(\varphi=0)$ quadrature.
    We see that quadrature squeezing can be observed in different intervals of the parameter range for all considered numbers of atoms.
    In particular, the strongest squeezing ($\approx 4\,\mathrm{dB}$ of squeezing, with $\braket{(\Delta \hat x)^2}_\textrm{vac}=1$) can be observed in the interval  $0.7 < r < 1.3$ for $N=5$.
    It is worth mentioning that this parameter region features high success probabilities; cf. Fig.~\ref{fig:success}.
    Additionally, increasing the number of atoms only increases the maximal value of squeezing marginally.
    However, the location of the maximal squeezing is decreasing with increasing $N$ which might be of interest in certain experimental scenarios.
    The strongest anti-squeezing is observed around $r=0.51$ and $1.5$.
    This corresponds to the cases in which the quantum state is broadly distributed in phase-space and its Wigner function shows pronounced negativities as we will see in Sec. \ref{sec:Wigner}.
    
\subsection{Sub-Poissonian light and Mandel $Q_{\mathrm{M}}$ parameter}
\label{sec:Mandel}
    We will now analyze the photon-number statistics of the generated cavity-field state.
    The photon-number statistics $c_n$ is obtained through the projection of the quantum state on the Fock basis
    \begin{align}
        c_n = \langle n|\hat \rho_{\mathrm{ps}}|n\rangle= \dfrac{1}{\mathrm{Tr}\left[\hat{\rho}^{N}_{22}\right]}  \dfrac{\left|\alpha\right|^{2n}}{n!}\cos^{2N}\left(r \sqrt{n}\right).
        \label{eq:PhotNumStat}
    \end{align}
    
    In particular, we are interested in the parameter regime where the PS state shows the nonclassical feature of sub-Poissonian light, i.e., having a photon-number distribution which is narrower than a Poissonian one.
    This characteristic can be identified via the Mandel $Q_{\mathrm{M}}$ parameter \cite{mandel_1979}, which is defined as
    \begin{align*}
        Q_{\mathrm{M}} &= \frac{\langle\left( \Delta\hat{n} \right)^2\rangle}{\langle\hat{n}\rangle} - 1 = \frac{\langle \hat{n}^2 \rangle - \langle \hat{n} \rangle^2}{\langle\hat{n}\rangle} - 1 ,
    \end{align*}
    where $\hat n$ is the photon-number operator.
    Sub-Poissonian light is indicated through $Q_{\mathrm{M}}<0$.
    
    To calculate the Mandel parameter for the conditional cavity-filed state in Eq. \eqref{eq:statePSN}, we evaluate the first two moments of the photon-number operator.
    The mean photon number is given by
    \begin{align*}
        \braket{\hat n} = \left|\alpha\right|^2 \dfrac{\beta(1)}{\beta(0)},
    \end{align*}
    and the second-order moment by
   \begin{align*}
        \langle\hat{n}^2\rangle = \dfrac{\left|\alpha\right|^2}{\beta(0)} \left(\left|\alpha\right|^2\beta(2) + \beta(1)\right),
    \end{align*}
    where we introduce the function $\beta(k)$  as
      \begin{equation*}
        \beta(k)=\sum_{l=0}^{\infty}\dfrac{\left|\alpha\right|^{2l}}{l!}\cos^{2N}\left(r \sqrt{l+k}\right).
      \end{equation*}
    Therefore, the Mandel $Q_{\mathrm{M}}$ parameter of the conditional cavity-field state is
     \begin{align*}
        Q_{\mathrm{M}} &= \left|\alpha\right|^2 \left(\frac{\beta(2)}{\beta(1)} - \frac{\beta(1)}{\beta(0)}\right).
    \end{align*} 
    
\begin{figure}[t]
        \center
        \includegraphics[width =  \columnwidth]{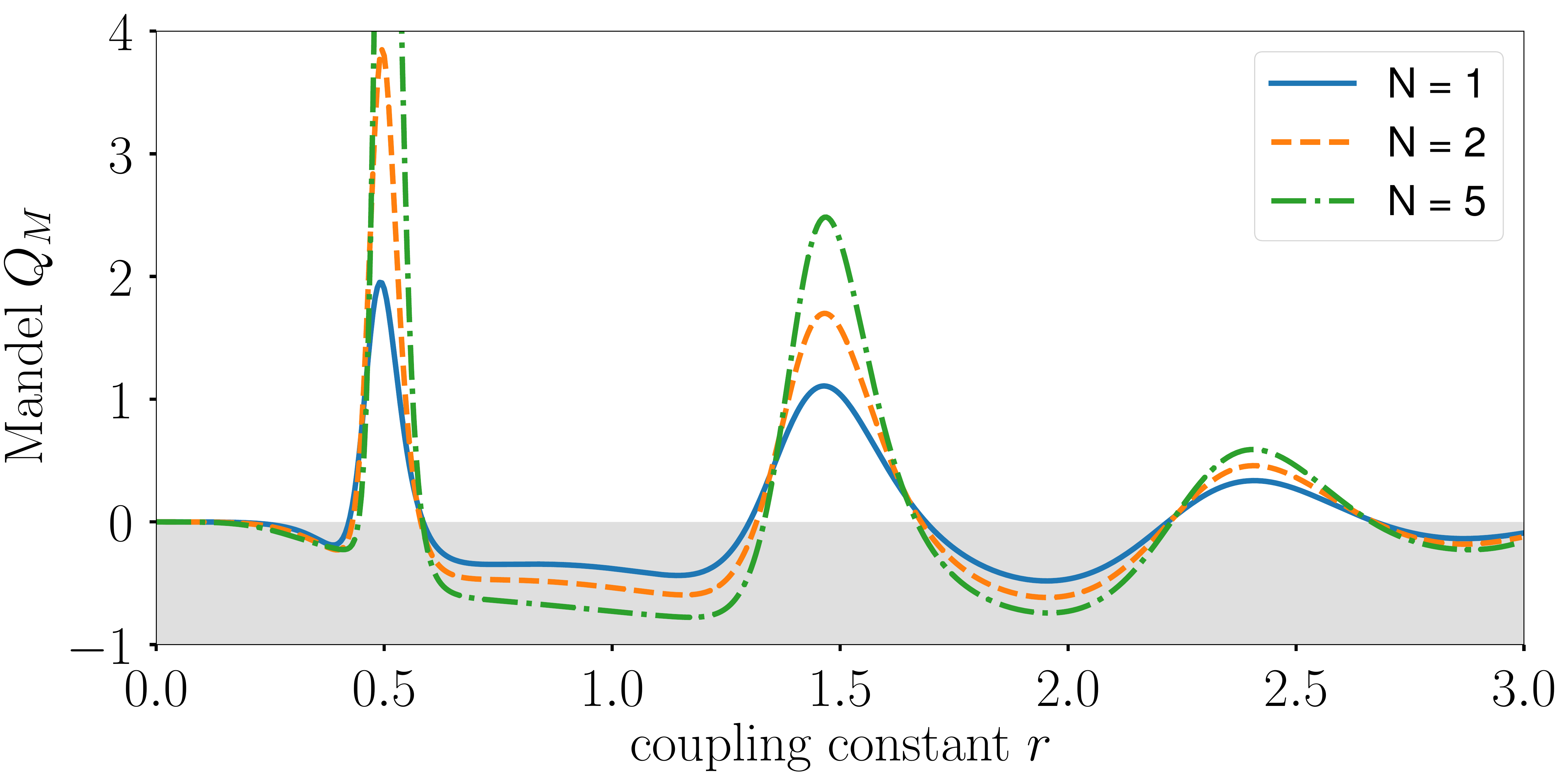} \caption{
        The Mandel $Q_{\mathrm{M}}$ parameter is shown for different numbers of PS atoms depending on the coupling parameter $r$.
        The states with sub-Poissonian photon-number distribution are characterized by parameters with negative $Q_{\mathrm{M}}$ (gray shaded area).
        }
        \label{fig:Mandel}
    \end{figure}

    The behavior of $Q_{\mathrm{M}}$ with respect to the coupling parameter $r$ and the number of PS atoms is shown in Fig. \ref{fig:Mandel}.
    We observe sub-Poissonian light $(Q_{\mathrm{M}}<0)$ for all considered numbers of PS atoms.
    In particular, in the intervals $0.7<r<1.3$ and $1.7<r<2.2$ the sub-Poissonian character of the photon-number distribution is clearly certified.
    The former interval concurs with a region for which we can also identify quadrature squeezing; cf. Fig \ref{fig:QuadSquee}.
    This is an interesting finding because the obtained quantum state shows these two important nonclassical features simultaneously.
    Typically, these two properties are studied independently which may lead to the false impression that they are mutually exclusive quantum attributes.
    Therefore, this case is particularly interesting, and may also serve as a didactic example of a state that possess squeezing and a sub-Poissonian photon-number statistic simultaneously.
    A similar behavior can also be observed for the parameter region around $r=2$.
    In this case, squeezing can, however, only be achieved by using one atom and the squeezing vanishes rapidly if one considers more atoms.
    
\subsection{Wigner function}
\label{sec:Wigner}
    The Wigner function links the density operator to a (quasi-)probability distribution in phase space and it encodes all information about the state of a given physical system \cite{wigner_1932}.
    It is a so-called quasiprobability distribution due to its possibility of attaining negative values. 
    These negative values are signatures of the nonclassicality of the state.
    
    \begin{figure*}[ht]
        \includegraphics[width=\textwidth]{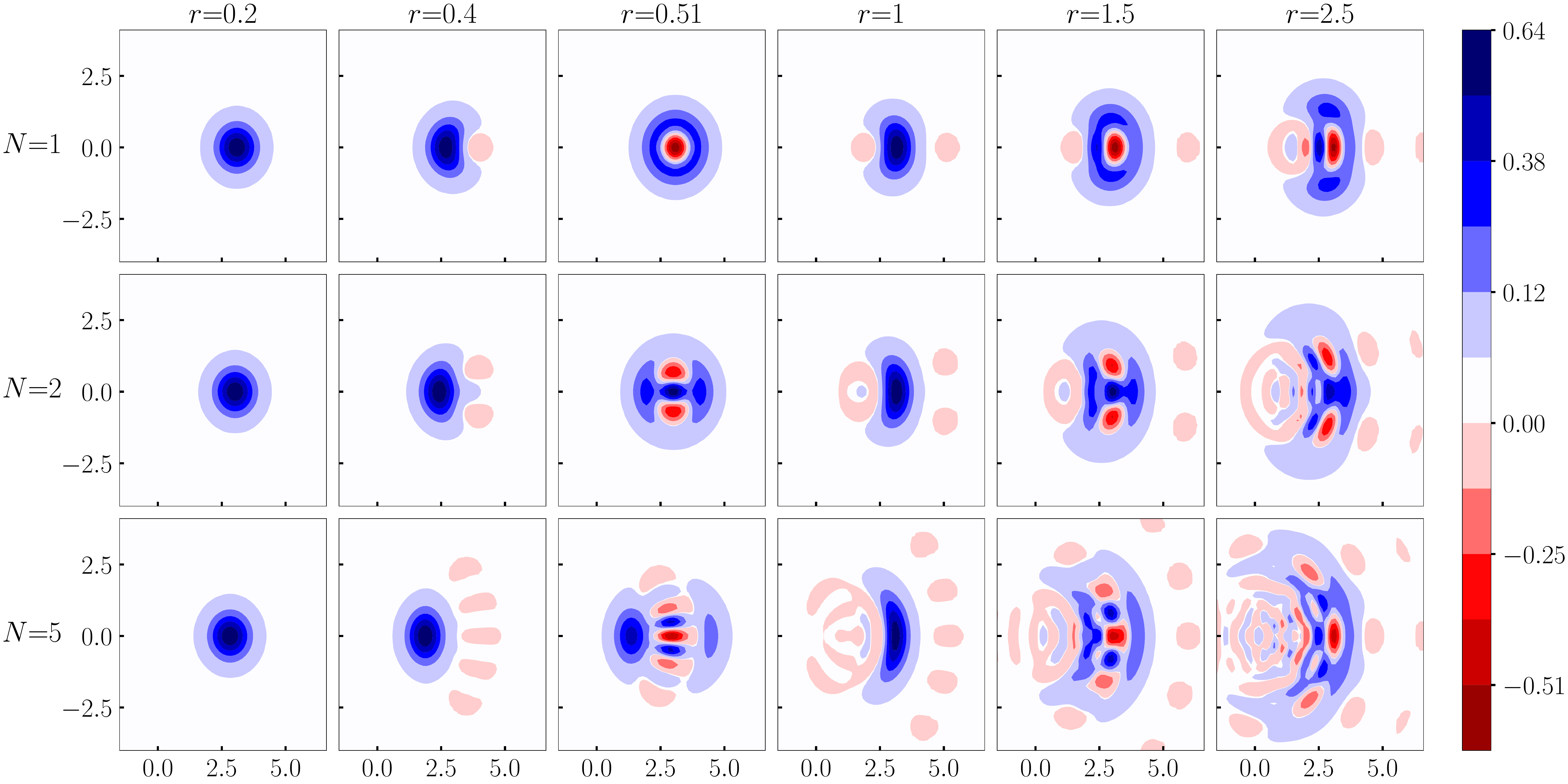}
        \caption{
        Wigner function of the conditional state.
        The rows correspond to the contour plot of the Wigner function in the phase space for different coupling parameters $r = 0.2,0.4,0.51,1,1.5,2.5$, while the columns indicate the different number of PS atoms, $N = 1,2,5$.
        The horizontal and the vertical axes represent the real and imaginary values of $\gamma$, respectively. 
        Red represents negative contributions, which clearly indicates the nonclassical character of the corresponding state.
        }
        \label{fig:Wigner}
    \end{figure*} 

    As initial condition, we consider the cavity field being prepared in a (classical) coherent state, which has a Gaussian Wigner function.
    We calculate the Wigner phase-space distribution of the PS cavity-field state and identify the parameter sets for which it attains negative values.
    In this way nonclassicality can be certified even if particular nonclassical features such as quadrature squeezing or a sub-Poissonian photon distribution cannot be observed.
    Although there exist various ways to calculate the Wigner function, in this work it proves most feasible to calculate the Wigner function using its relation to the Weyl equivalent
    \begin{align*}
        \mathrm{W}(\gamma) = \frac{2}{\pi^2} \mathrm{e}^{2|\gamma|^2} \int \; \mathrm{d}^2 \beta \braket{-\beta| \hat{\rho} |\beta} 
        \mathrm{e}^{2(\beta^*\gamma - \beta\gamma^*)}
    \end{align*}
    given in \cite{agarwal_1970}. 
    Inserting $\hat{\rho} = \hat{\rho}^{N}_{\mathrm{ps}}(r)$, it is possible to derive the Wigner function of the PS cavity-field state which yields
     \begin{align} \label{eq:Wigner}
        &\mathrm{W}(\gamma) = \dfrac{2\,\mathrm{e}^{-2|\gamma|^2-\left|\alpha\right|^2}}
        {\pi\mathrm{Tr}\left[\hat{\rho}^N_{22}\right]} 
        \Bigg(\sum_{n} \mathrm{c}_{nn} (\alpha,N)\,  (-1)^{n}\mathrm{L}^{0}_n\left(4|\gamma|^2\right) \\ \nonumber
        &+ \sum_{n > m}\mathrm{C}_{nm}(\gamma,\alpha,N)\, (-1)^{m} \sqrt{\frac{m!}{n!}}\frac{2^n}{2^m}  \mathrm{L}^{n-m}_m\left(4|\gamma|^2\right) \Bigg), 
    \end{align}
    with $\mathrm{C}_{nm}(\gamma,\alpha,N){=} 2|\gamma|^{m-n}|\alpha|^{m+n}\cos^N(r\sqrt{n}) \cos^N(r\sqrt{m})$ $\cos[\arg(\gamma)-\arg(\alpha)]/\sqrt{n!m!}$.
    For details on the calculation we refer to the Appendix~\ref{sec:appendix}.
    The resulting Wigner function is symmetric along the direction of the original displacement of the cavity field (the real axis for the states in Fig. \ref{fig:Wigner}).
    We can easily see this, since the only phase-dependent term in the Wigner function is the cosine of the relative phase $\arg(\gamma)-\arg(\alpha)$.
    In Fig.~\ref{fig:Wigner}, we show the Wigner function of the $N$-atom PS  cavity mode, for one, two, and five atoms and the interaction parameter $r$ varying from $0.2$ to $2.5$. 
    We observe in a wide parameter range that the Wigner functions of the obtained cavity-field states show negativities.
    This is a clear indicator of the nonclassical character of the quantum state.
    Let us study the behavior of the Wigner functions in more detail.
    First, consider the case with one PS atom, i.e., the first row in Fig.~\ref{fig:Wigner}.
    For $r = 0.2$ the deviation of the Wigner function from its original Gaussian form is still negligibly small and shows no visible nonclassical effect.
    With increasing $r$ the phase-space distribution is distorted around a negative region.
    The negativities are located first at the right side of the Wigner function ($r = 0.4$) and then at the center ($r = 0.51$) along the real axis of the phase space.
    While the negativity of the Wigner function is located closer at its center ($\alpha = \sqrt{10}\approx 3.2$) it becomes more negative. 
    With further increasing $r$ the Wigner function takes negative values in more regions, all along the real axes of the phase space.
    Overall it appears as if the negative regions move along the real axis from positive towards negative values.
    We observe similar behaviors for the Wigner functions for $N = 2$ and $N = 5$ corresponding to the second and third rows in Fig.~\ref{fig:Wigner}, respectively.
    The Wigner functions are also distorted and show negative regions, which are symmetrically distributed around the real axis.
    They differ in the number of negative regions, which are proportional to the number of atoms in the PS protocol. 
    In addition, the states with the lowest success probabilities (c.f. Fig.~\ref{fig:success}; $r = 0.51,1.5,2.5$), coincide with the states with the most negative values in the Wigner function. 
    Note that the states with $r = 0.51$ show the same symmetries as displaced cat states; cf, e.g., \cite{toscano_2006}. 
    In this sense theses states are interesting, however, they are rather unlikely to obtain due to their low success probabilities.
    Their success probabilities are $\approx 6.38\%$, $1.14\%$, and $0.06\%$  for $N=1,2,$ and $5$ atoms.
    Furthermore, it is important to mention that the observation of negativities in the Wigner function is a clear feature of non-Gaussian nonclassical states.
    Such states are highly interesting in the context of quantum information applications as they cannot be effectively simulated by classical computers \cite{veitch_2012,mari_2013}.

\section{Discussion}
\label{sec:discussion}

    For the preparation of quantum states with specific nonclassical features, it is important to identify the parameter regimes for which these properties can be achieved.
    We identify the regions of the parameter space where the cavity mode attains the different nonclassical behaviours.
    The regions with negative Mandel $Q$ parameter and negative Wigner function barely change with the number of atoms (c.f. Fig.~\ref{fig:Mandel} and~\ref{fig:Wigner}).
    In contrast, the appearance of squeezing is more dependent on the number of post selected atoms (c.f. Fig.~\ref{fig:QuadSquee}).  
    
    We perceive, that the cavity mode shows no squeezing or negative Mandel $Q_{\mathrm{M}}$ close to $r = 0.51, 1.5,  2.5$. 
    The oscillatory behavior of these quantities is related to the atomic Rabi oscillations. 
    In contrast, the Wigner function shows negativities for $r \gtrsim 0.3$. 
    Furthermore, we observe that squeezing is the hardest feature to be achieved, in the sense that the parameter regions for which we can observe it are the smallest.
    Unlike for squeezing, the parameter regions which feature a negative Mandel $Q_{\mathrm{M}}$ parameter are larger and especially also exist for higher values of $r$, and, therefore, have a larger overlap with regions with negative Wigner function.
    
    Let us also comment on the interpretation of the introduced protocol.
    At first glance, it seams surprising that the atoms entering and leaving the cavity in the ground state lead to a nonclassical field state, as no photons are added to or subtracted from the cavity field.
    They do, however, interact with the initial coherent state of the cavity field which features a Poissonian photon-number distribution.
    Importantly, for a fixed coupling parameter each photon-number contribution interacts differently with the passing atoms.
    Therefore, by the PS on the events where the atoms stay in the ground state certain photon-number contributions are more probable than others which leads to a redistribution of the photon-number statistics.
    Eventually, this leads to the observed nonclassical features in the cavity mode.
    
    We point out that the typical Jaynes-Cummings  interaction (one atom, no post-selection) already leads to nonclassical features of the cavity-mode and atom-field entanglement; c.f. \cite{raimond_2001,haroche_2006,walther_2006}.
    The single-mode nonclassical field properties are, however, rather weak for a wide parameter range.
    In particular, no significant squeezing \cite{meystre_1982} or sub-Poissonian photon-number statistics \cite{hillery_1987} can be observed in this case.
    Furthermore, a cat state of the cavity field \cite{haroche_2006} can only be reach for relatively high coupling parameters ($r\approx10$ for $\alpha=\sqrt{10}$).
    In addition, the introduced conditional-state preparation protocol provides the possibility to generate states with pronounced nonclassicality features for comparably low coupling parameters ($r<1$).
    
    Note that a similar system and post-selection approach was studied in \cite{nodurft_2019} in the context of attenuation without absorption.
    There, the authors discussed the possibility of attenuating and amplifying a coherent state of a traveling light field through the interaction and post-selection with several atoms.
    By studying the Husimi $Q$ phase-space distribution and the photon-number distribution of the light field for certain coupling parameters they argue that such a conditional-state preparation method allows to attenuate or amplify the field.
    Our analysis, however, shows that the action of the post-selection procedure does not only shift the overall amplitude of state, but severely changes the characteristics of the quantum state introducing various kinds of nonclassical properties.

\section{Conclusions}
\label{sec:conclusions}
    We introduce a method for the conditional preparation of nonclassical states of a cavity field.
    The studied system consists of two-level atoms passing through a high-quality cavity which interact with the cavity field in a resonant way. 
    The atoms are initially prepared in the ground state and the cavity field is initialized in a coherent state.
    The preparation of a nonclassical cavity-field state is achieved through conditioning on the cases in which the atom is detected also in the ground state after passing the cavity. 
    
    We calculate the conditional cavity-field state in the photon-number basis.
    The obtained state depends on the coupling constant, on the atom-light interaction, and the number of considered post-selection atoms.
    By controlling the coupling parameter, e.g., through control of the effective interaction time through a Stark switch, nonclassical states with different quantum characteristics can be obtained.
    In particular, we study quadrature squeezing, sub-Poissonian photon-number distributions and provided analytical expressions for the squeezing and the Mandel parameter in dependence on the coupling parameter and the number of passing atoms.
    Furthermore, the Wigner function of the cavity-mode was calculated and we could infer the state's nonclassical character from its negativities.
    For the preparation of states with different nonclassical features optimal parameter regions were identified.
    Summing up, the presented approach can easily be implemented in current cavity QED experiments and provides a versatile method for the engineering of nonclassical states of cavity fields.

\section*{Acknowledgements}
    KW thanks the European Commission of support through the Erasmus+ Traineeship Program.
    EA acknowledges funding from the European Union's Horizon 2020 research and innovation program under the Marie Sk\l{}odowska-Curie IF (InDiQE - EU Project No. 845486).
    MB acknowledges financial support by the Leopoldina Fellowship Program of the German National Academy of Science.

\appendix
\section{Derivation of Wigner function}
\label{sec:appendix}
    Here we show how the Wigner function of the post-selected state \eqref{eq:statePSN} is derived.
    The Wigner function can be calculated via \cite{agarwal_1970}
    \begin{align*}
        \mathrm{W}(\gamma) = \frac{2}{\pi^2}\, \mathrm{e}^{2|\gamma|^2} \int \, \mathrm{d}^2 \beta \braket{-\beta| \hat{\rho}^{N}_{\mathrm{ps}}(r) |\beta} 
        \mathrm{e}^{2(\beta^*\gamma - \beta\gamma^*)}.
    \end{align*}
    In a first step we calculate $\braket{-\beta| \hat{\rho}^{N}_{\mathrm{ps}}(r) |\beta}$ obtaining
    \begin{widetext}
    \begin{equation}
        \mathrm{W}(\gamma) = \frac{2}{\pi^2} \mathrm{e}^{2|\gamma|^2}\dfrac{\mathrm{e}^{-\left|\alpha\right|^2}}{\mathrm{Tr}[\hat{\rho}_{22}]} 
        \sum_{n,m}  \mathrm{c}_{nm} (\alpha,N) \frac{(-1)^{n}}{\sqrt{n!m!}} 
        \underbrace{\int \; \mathrm{d}^2 \beta \; \mathrm{e}^{-|\beta|^2} \mathrm{e}^{2(\beta^*\gamma - \beta\gamma^*)} \beta^m(\beta)^{*n}.}_{I_W}
        \label{eq:ApWigner}
    \end{equation}
    \end{widetext}
    
    The integral $I_W$ can be solved using derivative relations of the Fourier transformation
    \begin{align*}
        I_\mathrm{W} = \frac{(-1)^m}{2^{m+n}}\frac{\partial^n}{\partial\gamma^n}\frac{\partial^m}{\partial\gamma^{*m}} \pi \mathrm{e}^{-4\gamma\gamma^*}
    \end{align*}
    At this point the order of applying the derivatives becomes important. 
    To apply first $\frac{\partial^n}{\partial\gamma^n}$ corresponds to $n {\geq} m$ and applying first $\frac{\partial^m}{\partial\gamma^{*m}}$ to $m {\geq} n$.
    In the following, $I_W$ is split in two parts corresponding to the two cases given above and then reordered to fit the Rodrigues formula of the generalized Laguerre polynomial
    \begin{align*}
        L^\alpha_n(x) = \frac{x^{-\alpha}\mathrm{e}^x}{n!} \frac{\partial^n}{\partial x^n} \left(\mathrm{e}^{-x} x^{n+\alpha}\right).
    \end{align*}
    Defining $I_W^{(1)}$ for $m \geq n$ and $I_W^{(2)}$ for $n > m$ as
     \begin{equation}
        I_\mathrm{W}^{(1)} = \frac{2^m}{2^{n}} \pi \; n! \; \mathrm{e}^{-4|\gamma|^2} \gamma^{m-n} L_n^{m-n}(4|\gamma|^2)
        \label{eq:ApInt1}
    \end{equation}
    and
   \begin{align}
         \label{eq:ApInt2}
       I_\mathrm{W}^{(2)} = (-1)^{m-n} \frac{2^n}{2^{m}} \pi \; m! \; \mathrm{e}^{-4|\gamma|^2} \gamma^{*(n-m)}  L^{n-m}_m\left(4|\gamma|^2\right),     
   \end{align}
   respectively.
   Note that for $n = m$ both parts of $I_W$ are equal.
   Plugging Eqs. \eqref{eq:ApInt1} and \eqref{eq:ApInt2} into Eq. \eqref{eq:ApWigner} leads to
      \begin{widetext}
   \begin{align*}
        \mathrm{W}(\gamma) = \dfrac{2\,\mathrm{e}^{-2|\gamma|^2-\left|\alpha\right|^2}}
        {\pi\mathrm{Tr}\left[\hat{\rho}^N_{22}\right]} &\Bigg(\sum_{m\geq n} \mathrm{c}_{nm} (\alpha,N)\,  (-1)^{n} \sqrt{\frac{n!}{m!}}\frac{2^m}{2^n} \gamma^{m-n} \mathrm{L}^{m-n}_n\left(4|\gamma|^2\right) \\ 
        &+ \sum_{n > m} \mathrm{c}_{nm} (\alpha,N)\, (-1)^{m} \sqrt{\frac{m!}{n!}}\frac{2^n}{2^m} \gamma^{*(n-m)} \mathrm{L}^{n-m}_m\left(4|\gamma|^2\right) \Bigg) \\
     = \dfrac{2\,\mathrm{e}^{-2|\gamma|^2-\left|\alpha\right|^2}}
        {\pi\mathrm{Tr}\left[\hat{\rho}^N_{22}\right]} 
        &\Bigg(\sum_{n} \mathrm{c}_{nn} (\alpha,N)\,  (-1)^{n}\mathrm{L}^{0}_n\left(4|\gamma|^2\right) \\ \nonumber
        &+ \sum_{n > m}\mathrm{C}_{nm}(\gamma,\alpha,N)\, (-1)^{m} \sqrt{\frac{m!}{n!}}\frac{2^n}{2^m}  \mathrm{L}^{n-m}_m\left(4|\gamma|^2\right) \Bigg), 
    \end{align*}
    with 
    \begin{align*}
        \mathrm{C}_{nm}(\gamma,\alpha,N) &= 2\,\text{Re}[\mathrm{c}_{nm}(\alpha,N)\gamma^{(n-m)}]\\
        &= \frac{2|\gamma|^{m-n}|\alpha|^{m+n}\cos^N\left(r \sqrt{n}\right) \cos^N\left(r \sqrt{m} \right)}{\sqrt{n!m!}}\,\cos[\arg(\gamma)-\arg(\alpha)].
    \end{align*}
        \end{widetext}
\bibliography{biblio}
\end{document}